\documentclass[a4paper]{article}

\usepackage{INTERSPEECH2021}

\usepackage{anyfontsize}
\usepackage{t1enc}

\usepackage{graphicx}
\usepackage{amssymb,amsmath,bm}
\usepackage{textcomp}
\usepackage{color}
\usepackage{multirow}
\usepackage{hyperref}
\usepackage{url}
\usepackage[activate]{microtype}
\usepackage{subcaption}
\usepackage{soul}
\usepackage{cleveref}
\usepackage[activate]{microtype}
\usepackage{ulem}
\usepackage[nolist]{acronym}

\newcommand{\ie}{i.\,e., }
\newcommand{\eg}{e.\,g., }
\newcommand{\etal}{et\,al.\ }
\newcommand{\etalns}{et\,al.}
\newcommand{\ds}{\mbox{\textsc{Deep Spectrum }}}

\newcommand{\dsl}{\mbox{\textsc{DeepSpectrumLite }}}
\newcommand{\dslns}{\mbox{\textsc{DeepSpectrumLite}}}

\newcommand{\tfl}{\mbox{\textsc{TensorFlow Lite }}}

\def\vec#1{\ensuremath{\bm{{#1}}}}

\PassOptionsToPackage{dvipsnames,table}{xcolor}
\usepackage{pgfplotstable}
\pgfplotsset{compat=1.14}

\ninept

\title{DeepSpectrumLite: A Power-Efficient Transfer Learning Framework for Embedded Speech and Audio Processing from Decentralised Data}
\name{Shahin Amiriparian$^1$, Tobias H\"ubner$^1$, Maurice Gerczuk$^1$, Sandra Ottl$^1$, Bj\"orn W.\ Schuller$^{1,2}$}

\address{\fontsize{11}{11}\selectfont 
  $^1$EIHW -- Chair of Embedded Intelligence for Health Care and Wellbeing, University of Augsburg, Germany\\
  $^2$ GLAM -- Group on Language, Audio, \& Music, Imperial College London, UK}
\email{shahin.amiriparian@uni-a.de, tobias.huebner@informatik.uni-augsburg.de,  maurice.gerczuk@informatik.uni-augsburg.de, sandra.ottl@informatik.uni-augsburg.de, schuller@ieee.org}

\begin{document}

\maketitle
\begin{abstract}
Deep neural speech and audio processing systems have a large number of trainable parameters, a relatively complex architecture, and require a vast amount of training data and computational power. These constraints make it more challenging to integrate such systems into embedded devices and utilise them for real-time, real-world applications.  
We tackle these limitations by introducing \dslns, an open-source, light-weight transfer learning framework for on-device speech and audio recognition using pre-trained image convolutional neural networks (CNNs). The framework creates and augments Mel-spectrogram plots on-the-fly from raw audio signals which are then
used to finetune specific pre-trained CNNs for the target classification task. Subsequently, the whole pipeline can be run
in real-time with a mean inference lag of $242.0$\,ms when a \textsc{DenseNet121} model is used on a consumer grade \textit{Motorola moto e7 plus} smartphone. 
\dsl operates decentralised, eliminating the need for data upload for further processing. 
By obtaining state-of-the-art results on a set of paralinguistics tasks, we demonstrate the suitability of the proposed transfer learning approach for embedded audio signal processing, even when data is scarce. 
We provide an extensive command-line interface for users and developers which is comprehensively documented and publicly available at {\url{https://github.com/DeepSpectrum/DeepSpectrumLite}}.

\end{abstract}

\section{Introduction}
\label{sec:introduction}
Over the past decade, the number of wearable devices such as fitness trackers, smartphones, and smartwatches has increased remarkably~\cite{van2015curse}. 
With rising amount of sensors, these devices are capable of gathering a vast amount of users' personal information, such as state of health~\cite{ko2010wireless}, speech, or physiological signals including skin conductance, skin temperature, and heart rate\cite{schuller2013automatic}. 
In order to automatically process such data and obtain robust data-driven features, deep representation learning approaches~\cite{freitag2017audeep,amiriparian2019you} and end-to-end learning methodologies~\cite{tzirakis2018end2you} can be applied. These networks, however, have large number of trainable parameters (correlated with the large model size) and need high amount of data to achieve a good degree of generalisation~\cite{zhao2019object}. These factors increase the energy consumption of the trained models~\cite{yang2017method} and confine their real-time capability. Furthermore, whilst personal data in unprecedented volumes is `in transit' or being synchronised with the cloud for further processing, it is susceptible to eavesdropping~\cite{cilliers2020wearable} 
and this issue raises privacy and security concerns for the user (\eg discriminatory profiling, manipulative marketing)~\cite{montgomery2018health}. Such restrictions emerged the need for novel neural architectures and collaborative machine learning techniques without centralised training data~\cite{li2020federated}. Recent advancements include `squeezed' neural architectures~\cite{iandola2016squeezenet} and the federated learning paradigms~\cite{li2020federated}. 
Iandola~\etal have introduced \textsc{SqueezeNet}, a `pruned' \ac{CNN} architecture with $50\times$ less trainable parameters than \textsc{AlexNet}~\cite{krizhevsky2012imagenet} with an \textsc{AlexNet}-level accuracy~\cite{iandola2016squeezenet}. A more squeezed architecture, \textsc{SqueezeNext}, with $112\times$ less parameters than \textsc{AlexNet} (with similar accuracy) was introduced by Gholami~\etalns~\cite{gholami2018squeezenext}. In 2019, Mehta~\etalns~\cite{mehta2019espnetv2} have proposed \textsc{ESPNetv2}, a light-weight general purpose \ac{CNN} with point-wise and depth-wise dilated separable convolutions for representation learning from large receptive fields with fewer parameters. Further energy-efficient \ac{CNN} architectures have been proposed and applied for traffic sign classification~\cite{zhang2020lightweight} and optical flow estimation~\cite{hui2018liteflownet}.
\begin{figure*}[th!]
\begin{center}
\includegraphics[width=1\textwidth]{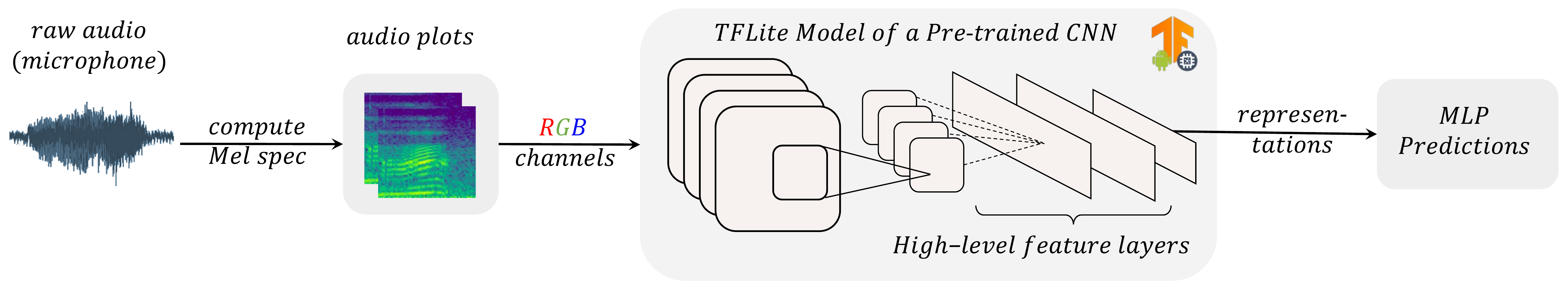}
\end{center}
\caption{A general overview of a \dsl model deployed on a target device for inference. Raw audio (from the device's microphone) is first converted to a spectrogram representation and the values are mapped to the red-green-blue (RGB) colour space according to a certain colour mapping definition. These spectrogram plots are then forwarded through the TFLite version of a trained \ac{CNN} model and a \ac{MLP} classifier head generates predictions for the task at hand.}
\label{fig:dsl}
\end{figure*}

For effective utilisation of deep \acp{CNN} and to cope with data scarcity in the field of audio signal processing, we have introduced the \ds system\footnote{\href{https://github.com/DeepSpectrum/DeepSpectrum}{https://github.com/DeepSpectrum/DeepSpectrum}}~\cite{amiriparian2017Snore} at INTERSPEECH 2017. In~\cite{amiriparian2017Snore}, we have forwarded (Mel-)spectrogram plots of audio signals with different colour mappings  through pre-trained \acp{CNN} and extracted the activations of the penultimate fully connected layer of these networks as feature set. For the effect of different colour maps on the representations, please refer to~\cite{amiriparian2017Snore,amiriparian2019you,amiriparian2020towards}. \ds features have shown to be effective for a variety of paralinguistics and general audio recognition tasks, including speech emotion recognition (SER)~\cite{ottl2020group}, sentiment analysis~\cite{amiriparian2017sentiment}, and acoustic surveillance~\cite{amiriparian2018bag}. Furthermore, the \ds system has demonstrated to be a competitive baseline system for the 2018--2021 editions of the \acf{ComParE}~\cite{schuller2018interspeech,schuller2019interspeech,schuller2020interspeech,Schuller21-TI2}.  
In this paper, we propose \dslns, an extension of the \ds framework for embedded speech and audio processing. Whereas the \ds framework extracts features from pre-trained \acp{CNN}, \dsl goes one step ahead. First, the \dsl framework adds a light-weight \acf{MLP} to the neural network pipeline. The added \ac{MLP} is responsible for either the classification or for regression. Second, the \dsl frameworks offers support for efficient on-device computation of the audio signal processing including the generation of spectrogram image plots. The proposed \dsl system implements a model and inference structure that is focused on mobile usage. We make our \dsl framework publicly available for users and developers on GitHub\footnote{\href{https://github.com/DeepSpectrum/DeepSpectrumLite}{https://github.com/DeepSpectrum/DeepSpectrumLite}} and PyPI.

\section{Proposed System}
\label{sec:system}
Our framework is composed of two main parts: i) task-specific, transfer learning-based model training (cf.~\Cref{ssec:model_training}), and ii) decentralised audio processing using the trained model (cf.~\Cref{ssec:integration_into_device}).

\subsection{Task-specific Transfer Learning}
\label{ssec:model_training}

The input of our system are raw audio signals with a sample rate of $16$\,kHz. For simplicity, our system reads only one audio channel. Subsequently, we apply a sliding window function in order to split the audio signals into smaller fixed-width chunks. For each chunk, we apply a signal normalisation and then a \ac{STFT} with Hanning windows of $32$\,ms and $50.0$\,\% hop length. The spectrograms are then transformed to Mel spectrograms with $128$ Mel bins. We further compute the power spectral density on the $dB$ power scale, and apply a \textit{min-max normalisation} which is linearly scaled between \([0, 255]\). Subsequently, each value in the rescaled spectrogram matrix is mapped according to the \textit{viridis} colour definition. Because we use Image \acp{CNN} that were pre-trained on ImageNet~\cite{huang2017densely,deng2009imagenet}, we resize the spectrogram image plot to $224\times224$ pixels with bi-linear interpolation and mean normalise the image colour channel values according to the original ImageNet dataset.
Afterwards, we use the deep \ac{CNN} model \textsc{DenseNet121}~\cite{huang2017densely} as a convolutional feature extractor for the generated audio plot images and attach an \ac{MLP} classifier containing a single hidden layer with Attention-based-Rectified-Linear-Unit (AReLU)~\cite{chen2020arelu} activation on top of this base. In order to reduce the effect of overfitting, we further apply the regularisation technique Dropout.
The training of our transfer learning models then proceeds in two phases.
In the first phase, we freeze the \ac{CNN} model structure and only train the classifier head. In the second phase, we unfreeze a part of the \ac{CNN}'s layers and continue training with a reduced learning rate. 

Furthermore, we apply different data augmentation techniques to the spectrogram plots on-the-fly during training. Data augmentation helps to reduce the effect of overfitting, especially when only a small number of training samples is available. 
\dsl has implemented an adapted version of the SapAugment data augmentation policy~\cite{hu2021sapaugment}. The policy decides for every training sample its portion of applied data augmentation. We apply both CutMix~\cite{yun2019cutmix} and SpecAugment~\cite{Park_2019} data augmentations relatively to the loss value of all samples within a batch. The basic idea of SapAugment is that a training sample with a comparably low loss value is easy to understand using the current weights of a neural network, therefore, more data augmentation can be applied. Whereas, when a sample has a comparably high loss, SapAugment argues that less data augmentation should be applied until the sample reaches a low loss value. For details how the portion of applied data augmentation relative to the loss value is computed, we refer to~\cite{yun2019cutmix}.

\subsection{Decentralised Audio Processing}
\label{ssec:integration_into_device}
After a task-specific model is trained, its network structure and weights are saved into a \ac{HDF} version 5. The saved model is then converted to \textsc{TensorFlow} (TF) Lite\acused{TF}\footnote{\href{https://www.tensorflow.org/lite}{https://www.tensorflow.org/lite}} format for compatibility on embedded devices. 
Since our framework applies all necessary preprocessing steps within the data pipeline structure, there is no device-specific implementation required.
A schematic overview of \dsl deployed on a target mobile device is depicted in~\Cref{fig:dsl}. From the input raw audio signals (\eg signals captured from a microphone) Mel spectrogram plots are created which are then forwarded through a \tfl version of the model trained as described in~\Cref{ssec:model_training}. It consists of a (fine-tuned) image \ac{CNN}, here a \textsc{DenseNet121}, and a light-weight \ac{MLP} head which classifies the deep representations obtained from a specific layer of the \ac{CNN}. 
\section{Experiments}
\label{sec:experiments}
The evaluation of our proposed system is two fold. First, we perform experiments regarding the general learning capabilities of \dsl by comparing its efficacy on four databases against the more traditional \ds feature extraction pipeline utilising a linear \ac{SVM} as classifier. Second, we investigate the suitability of a trained \dsl model for real-time audio classification on an embedded device.

\subsection{Datasets}
\label{ssec:datasets}
\begin{table}
\footnotesize
\centering
\caption{Statistics of the databases utilised in our experiments in terms of number of speakers (\emph{Sp.}), the number of classes (\emph{C.}), and the total duration (\emph{Dur.}).} 
\label{tab:dataset-stats}
\resizebox{\columnwidth}{!}{
  \renewcommand*{\arraystretch}{1.4}
\begin{tabular}[t]{lrrrr}
    \toprule
    \textbf{Name} &  \textbf{\#} & \textbf{C.} & \textbf{Sp.}  & \textbf{Dur. [h]} \\ 
    \midrule
    \textbf{\acs{CCS}}: COVID-19 Cough & 725 & 2 & 397 & 1.63 \\
    \textbf{\acs{CSS}}: COVID-19 Speech & 893 & 2 & 366 & 3.24  \\ 
    \textbf{\acs{ESS}}: Escalation in Speech & 914 & 3 & 21 & 0.97  \\ 
    \textbf{\acs{IEMOCAP}}: Emotional Speech &  5\,531 & 4 & 10 & 7.0 \\ 
    \bottomrule
\end{tabular}
}
\end{table}
We apply a wide range of tests on four different corpora. All datasets are speaker-independently split into training, validation, and test partitions. 
The \ac{IEMOCAP} dataset~\cite{busso2008iemocap} is an English emotion dataset containing audio of (scripted and improvised) dialogues between $5$ female and $5$ male speakers, adding up to $5\,531$ utterances. The chosen emotion classes are happiness (fused with excitement), sadness, anger, and neutral. The dataset is split into session $1$ to $3$ for training, session $4$ for validation, and session $5$ for testing.
Furthermore, we test the \dsl framework with the \ac{CCS} and the multi-language \ac{CSS} corpora which are both part of this year's \ac{ComParE} Challenge~\cite{Schuller21-TI2}. \ac{CCS} consists of crowd-sourced audio samples of coughing, recorded from $397$ subjects resulting in $725$ clips. \ac{CSS} contains $893$ audio samples from $366$ subjects.
A preceding COVID-19 test of the subjects was \textit{positive} for one part, and \textit{negative} for the rest. The result of this test should be predicted by the challenge participants based on the audio content.
Also part of this year's \ac{ComParE} Challenge is the \ac{ESS} corpus, combining the dataset of aggression in trains~\cite{Lefter2013} and the stress at service desk dataset~\cite{Lefter2014}. In total, $21$ subjects were exposed to different scenarios that were recorded in $914$ audio files. The original labels are mapped onto a 3-point scale, \textit{low}, \textit{medium}, and \textit{high} escalation. The language in the clips is Dutch. 
For further information about the \ac{CCS}, \ac{CSS}, and \ac{ESS} datasets, the reader is referred to the \ac{ComParE} Challenge baseline paper~\cite{Schuller21-TI2}.

\subsection{Hyperparameters}
\begin{table}[t!]
 	\caption{This table shows the configuration of the different hyperparameters for each of the datasets used in our experiments.
 	}
	\label{tab:hyperparameters}
\centering
  \resizebox{1.0\columnwidth}{!}{
  \renewcommand*{\arraystretch}{1.3}
\begin{tabular}{lrrrr}
				\toprule
				\textbf{Hyperparameter} & \textbf{CCS} & \textbf{CSS} & \textbf{ESS} & \textbf{IEMOCAP} \\ \midrule
				Classifier units   &  512 & 700 & 512 & 512                \\
				Dropout rate   &  0.25 & 0.4 & 0.25 & 0.25                \\ 
				Initial learning rate   &  0.001 & 0.01 & 0.001 & 0.001      \\ 
				Epochs of first phase  &  40 & 20 & 40 & 40      \\
				Epochs of second phase  &  200 & -- & 200 & 200     \\
				Fine-tuned layers  &  298 & 0 & 298 & 128     \\
				Audio chunk length [s]   &  3.0 & -- & 3.0 & 4.0              \\  \bottomrule
			\end{tabular}%
			}
\vspace{-0.5cm}
\end{table}

We train our models with the AdaDelta optimiser on the cross entropy loss function in batches of $32$ samples. After training the classifier head for a certain number of initial epochs only, we reduce the learning rate $10$ fold and continue training with some of the layers of the \textsc{DenseNet121} unfrozen. Because the datasets have different sizes and number of classes, we slightly adapt our hyperparameter configuration to each of them. We refer the reader to~\Cref{tab:hyperparameters} for more details. Furthermore, we evaluate four data augmentation configurations: 1) no augmentation, 2) only CutMix, 3) only SpecAugment, and finally 4) both CutMix and SpecAugment.
In our experiments, we use SapAugment with the configuration values \(a=0.5, s=10\). Our CutMix algorithm hyperparameters are set to cut and paste squared patches between a size of \([0\,px, 56\,px]\) among the training samples. The ground truth labels are proportionally mixed according to the pasted patch size. Moreover, the SpecAugment data augmentation creates one time mask and one frequency mask for every training sample. The size of every mask is between \([0.0\,px, 67\,px]\).
The actual patch sizes and mask sizes depend on the samples' loss value (cf.~\ref{ssec:model_training}). Because the number of available training samples is limited, we expect the problem of underfitting when applying data augmentation for every single training sample. Therefore, we throttle down the usage of all data augmentations by adding an execution probability between \([10.0\,\%, 25.0\,\%]\) dependent on the sample's loss value.
\subsection{Results}
\label{ssec:results}
We evaluate the performance on the test partitions using the \ac{UAR} metric because it gives more meaningful information when a dataset has an unbalanced number of samples in its classes. \Cref{tab:results} compares the \dsl models against the regular \ds feature extraction framework with an \ac{SVM} classifier. For comparison, the network configuration of \dsl is taken from this year's \ac{ComParE} Challenge~\cite{Schuller21-TI2}. Because the \ac{IEMOCAP} dataset is not part of the \ac{ComParE} Challenge, we apply the same experimental settings in order to reproduce the results. 
Furthermore, to be consistent with the \ac{ComParE} methodology, we first optimise our models for the validation partition and apply a test with the best configuration where the training and validation sets are fused together.
We additionally provide $95.0\,\%$ \acp{CI} on the test partitions. They were obtained by $1000 \times$ bootstrapping. In each iteration, a random selection of test samples is replaced and the \ac{UAR} is computed.

\begin{table*}[t!]
 	\caption{Results of the transfer learning experiments with \dsl (\textsc{DS Lite}) on three of this year's \ac{ComParE} Challenge tasks (\ac{CCS}, \ac{CSS}, and \ac{ESS}) and \ac{IEMOCAP} compared against \ds feature extraction + \ac{SVM}. For the \ac{ComParE} tasks, we evaluate against the official \ds results presented in~\cite{Schuller21-TI2} while for IEMOCAP, we run the \ds challenge baseline with the same settings ourselves.
 	\textbf{CCS}: COVID-19 Cough. \textbf{CSS}: COVID-19 Speech. \textbf{ESS}: Escalation in Speech.\break \textbf{IEMOCAP}: Emotional Speech. \textbf{UAR}: Unweighted average recall. \textbf{CI}: $95\%$ confidence interval.
 	}
	\label{tab:results}
\centering
  \resizebox{1.0\textwidth}{!}{
  \renewcommand*{\arraystretch}{1.3}
\begin{tabular}{lrrrrrrrrrrrr}
    \toprule
    & \multicolumn{3}{c}{\bf  \textsc{CCS}}  &  \multicolumn{3}{c}{\bf  \textsc{CSS}} &      \multicolumn{3}{c}{\bf  \textsc{ESS}} & \multicolumn{3}{c}{\bf  \textsc{IEMOCAP}}       \\    
    \relax
    [\acs{UAR} \%] &                    Dev  &      Test  &     CI on Test &    Dev  &      Test  &     CI on Test &    Dev  &  Test &  CI on Test &    Dev &   Test &  CI on Test  \\ \midrule 
    \ds + SVM~\cite{Schuller21-TI2} &   63.3 &      64.1 &      55.7-72.8  &    56.0 &      60.4 &      55.9-64.9 &     64.2  & 56.4  & 51.5-61.3 &     53.0  & 56.3  & 54.2-58.2 \\  
    
    \textsc{DS Lite} (no augmentation) &     56.5  &     71.1  &     62.2-79.5 &     61.6  &     61.2  &     55.1-66.8 &     43.1  & 60.0  & 54.3-66.1 &     55.1  & 56.4  & 55.2-63.9  \\   
    \textsc{DS Lite} (CutMix) &              57.1  &     71.4  &     62.5-79.4 &     62.2  &     62.3  &     56.4-68.4 &     43.1  & 59.9  & 54.1-65.7  &    55.5  & \textbf{59.7}  & 55.6-64.0 \\      
    \textsc{DS Lite} (SpecAugment) &         58.4  &     72.7 &      63.9-80.9  &     60.7  &     63.6  &     58.1-69.0 &     48.1  & 61.3  & 55.4-67.2 &     55.2  & 59.3  & 54.9-63.5  \\      
    \textsc{DS Lite} (CutMix+SpecAugment) &  59.0  &     \textbf{74.4}  &     66.3-82.4 &     60.7  &     \textbf{63.9}  &     55.1-66.8 &     47.2  & \textbf{61.7}  & 55.8-67.3  &     53.9  & 59.2  & 54.9-63.5 \\  
\bottomrule
\end{tabular}
}
\end{table*}

\subsection{Computational performance}
\label{ssec:computational_performance}
The number of \acp{FLOP} is a measure of the efficiency of a computer system or an algorithm. The more \acp{FLOP} an algorithm needs to finish, the longer it takes to run and the more power it consumes.
Embedded devices typically have a limited power capacity as they have a battery and no continuous power supply. Therefore, we take the \dsl framework's power efficiency into account. This subsection examines the models' \acp{FLOP}, mean execution time, mean of requested memory, and the model size. Our analysis is split into the audio signal preprocessing step, \ie the spectrogram plot creation, and the final model inference.
Both the \acf{TF} model, and the spectrogram creation were executed $50$ times on a $2.3$\,GHz Quad-Core Intel Core i5 CPU with $4$ threads.
In order to investigate the difference to the \ac{TF} Lite model, we tested both models on the same system. Furthermore, we examined an on-device test on a consumer grade smartphone \textit{Motorola moto e7 plus} which comes with a $4\times1.8$\,GHz Kryo 240, a $4\times1.6$\,GHz Kryo 240, and an Adreno 610 GPU. Every on-device test was repeated $50$ times. 
~\Cref{tab:performance} shows the performance results of our spectrogram image creation and the \textsc{DenseNet121} model which includes the classification layers as well. The spectrogram creation has a model size of $150.0$\,kb, a mean execution time of $7.1$\,ms, and it consumes $4.5$\,MB memory. Because the plot generation is not a \ac{TF} model, we cannot measure the \acp{FLOP} nor are there any parameters. However, the number of \acp{FLOP} is expected to be small based on the measured execution time. During the transformation from the \ac{TF} \ac{HDF} file format to the \ac{TF} Lite model, the model size is reduced by the factor of $2.7$.
Although the \ac{TF} Lite model consumes more memory than the regular \ac{TF} model, the mean inference time is reduced by $150.3$\,ms measured on the same CPU setup. 
The \ac{TF} Lite model has a mean inference time of $242.0$\,ms on our embedded device.
\begin{table}[t!]
 	\caption{This table shows the mean execution time, the number of parameters, the mean requested memory, and the model size of our preprocessing (prepr.), the \textsc{DenseNet121} \textsc{TensorFlow} (TF), and \ac{TF} Lite model. In the \ac{TF} Lite Model column, the values before the slash are from the test on the CPU system, whereas the values after the slash are from the on-device test. Details regarding the test setup are described in the text. \break \textbf{FLOPs}: Floating point operations.
 	}
	\label{tab:performance}
\centering
  \resizebox{1.0\columnwidth}{!}{
  \renewcommand*{\arraystretch}{1.3}
\begin{tabular}{lrrrr}
				\toprule
				\multicolumn{1}{l}{} & \textbf{Prepr.} & \textbf{TF Model} & \textbf{TF Lite Model} \\ \midrule
				Mean time {[}ms{]}   &  7.1 & 240.0 & 89.7 / 242.0                 \\
				FLOPs   &  -- & 3.1\,G & --                 \\ 
				Parameters   &  -- & 7.6\,M & 7.6 M       \\ 
				Mean memory {[}MB{]}  &  4.5 & 116.5 & 185.4 / 292.8     \\
				Model size  &  150.0\,kb & 82.1\,MB & 30.0\,MB     \\ \bottomrule
			\end{tabular}%
			}
\end{table}
\section{Discussion}
\label{sec:discussion}
The results achieved with \dsl (described in~\Cref{ssec:results}) on four exemplary paralinguistic analysis tasks show the system's efficacy, in particular compared to the traditional \ds feature extraction and \ac{SVM} pipeline which is consistently outperformed on the test partitions. Furthermore, the applied state-of-the-art data augmentation techniques (CutMix and SpecAugment) in combination with an adapted version of the SapAugment~\cite{hu2021sapaugment} policy proved themselves to be especially useful on the smaller datasets (\ac{CCS}, \ac{CSS}, and \ac{ESS}) from this year's \ac{ComParE} Challenge. On both \ac{CCS} and \ac{ESS}, data-augmentation helps to raise achieved test set \acp{UAR} above the official challenge baselines. 
For the \ac{IEMOCAP} dataset, our best performing model is comparable to other state-of-the-art audio-based speaker-independent approaches. 
Although we obtain outperforming results on the test partitions, the models do not yet outperform on the development partitions. This could be explained by the fact that the \ac{MLP} classifier requires many training samples. Since our test is based on a larger training dataset (fusion with validation), the \ac{MLP} has a sufficient amount of data. 
Considering embedded devices, such as consumer grade smartphones, as deployment targets, \dsl is further suitable for real-time speech recognition tasks. With a total inference time of only a quarter of a second for a three second long raw audio chunk, time continuous analysis from raw microphone input can be performed directly on-device. The measured performance, both in terms of recognition accuracies on the datasets as well as inference times, make \dsl a powerful framework for many paralinguistic recognition tasks where data is often scarce and sometimes of a sensitive nature.  
\section{Conclusion}
\label{sec:conclusions}

In this paper, we presented a framework for training and deploying power-efficient deep learning models for embedded speech and audio processing. By making use of transfer learning from ImageNet pre-trained deep \acp{CNN} with spectrogram inputs and state-of-the-art data-augmentation techniques, \dsl can produce powerful speech analysis models that can then be easily deployed to embedded devices as an end-to-end prediction pipeline from raw microphone input. 

For future work, further reductions in model size can be pursued. In this paper, we purposefully chose DenseNet121 as base model for our experiments to be able to directly compare our results to the official \ac{ComParE} Challenge baselines. However, from an efficiency standpoint of view, networks specifically designed with smaller memory and computation footprints in mind, such as SqueezeNet~\cite{iandola2016squeezenet} or SqueezeNext~\cite{gholami2018squeezenext}, can be a better choice for the targeted applications and thus should be evaluated as feature extractors in \dsl. Finally, techniques such as pruning and quantisation~\cite{han2015deep} can be explored together with their impacts on speed and model accuracy. 

\section{Acknowledgements}
This research was partially supported by Deutsche Forschungsgemeinschaft (DFG) under grant agreement No.\ 421613952 (ParaStiChaD), and Zentrales Innovationsprogramm Mittelstand (ZIM) under grant agreement No.\ 16KN069455 (KIRun).

\bibliographystyle{IEEEtran}

\bibliography{mybib}

\begin{acronym}
\acro{CNN}[CNN]{Convolutional Neural Network}
\acrodefplural{CNN}[CNNs]{Convolutional Neural Networks}
\acro{FLOP}[FLOP]{floating point operation}
\acrodefplural{FLOP}[FLOPs]{floating point operations}
\acro{MLP}[MLP]{multi-layer perceptron}
\acrodefplural{MLP}[MLPs]{multi-layer perceptrons}
\acro{HDF}[HDF]{Hierarchical Data Format}
\acro{STFT}[STFT]{Short-Time Fourier Transform}
\acrodefplural{STFT}[STFTs]{Short-Time Fourier Transforms}
\acro{SVM}[SVM]{Support Vector Machine}
\acro{IEMOCAP}[IEMOCAP]{Interactive Emotional Dyadic Motion Capture}
\acro{CI}[CI]{Confidence interval}
\acrodefplural{CI}[CIs]{Confidence intervals}
\acro{CCS}[CCS]{COVID-19 Cough}
\acro{CSS}[CSS]{COVID-19 Speech}
\acro{ESS}[ESS]{Escalation at Service-desks and in Trains}
\acro{UAR}[UAR]{Unweighted Average Recall}
\acrodefplural{UAR}[UARs]{Unweighted Average Recall}
\acro{TF}[TF]{TensorFlow}
\acro{ComParE}[ComParE]{Computational Paralinguistics Challenge}
\acrodefplural{ComParE}[ComParE]{Computational Paralinguistics Challenges}
\end{acronym}

\end{document}